\documentclass[12pt]{article}
\usepackage{amstex}
\newcommand{\PT}{\mbox{\bf P}}
\newcommand{\NP}{\mbox{\bf NP}}
\newcommand{\coNP}{\mbox{\bf co--NP}}
\newcommand{\BPP}{\mbox{\bf BPP}}
\newcommand{\PSPACE}{\mbox{\bf PSPACE}}
\newcommand{\BQP}{\mbox{\bf BQP}}
\newcommand{\QP}{\mbox{\bf QP}}
\newcommand{\ZQP}{\mbox{\bf ZQP}}
\newcommand{\EQP}{\mbox{\bf EQP}}
\newcommand{\ZPP}{\mbox{\bf ZPP}}
\newcommand{\BQPs}{\mbox{\scriptsize\bf BQP}}
\newcommand{\sharpPs}{\mbox{\scriptsize\#\bf P}}

\newtheorem{theorem}{Theorem}
\newtheorem{lemma}[theorem]{Lemma}
\newtheorem{corollary}[theorem]{Corollary}

\newenvironment{proof}{\begin{trivlist}\item[]{\flushleft\bf Proof }}
     {\hspace*{\fill}\raisebox{-1pt}{\boldmath$\Box$}\end{trivlist}}
\newenvironment{example}{\trivlist \item[\hskip \labelsep{\bf Example}]}
     {\hspace*{\fill}\raisebox{-1pt}{\boldmath$\Box$}}

\def\ket#1{\mbox{$| #1 \rangle$}}
\def\integer{\Bbb Z}             
\DeclareMathSymbol{\leqslant}{\mathrel}{AMSa}{"36}  
\def\subgroup{\leqslant}         

\title{On The Power of Exact Quantum Polynomial Time}\,%

\author{
Gilles Brassard\,%
\thanks{\,Supported in part by Canada's {\sc nserc} and Qu\'ebec's
{\sc fcar}.}\\
{\protect\small\sl Universit\'e de Montr\'eal\/}\,%
\thanks{\,D\'epartement IRO, Universit\'e de Montr\'eal,
C.P. 6128, succursale centre-ville, Montr\'eal (Qu\'ebec), 
Canada H3C 3J7.
email: {\tt brassard$\mathchar"40$iro.umontreal.ca}.}
\and
Peter H{\o}yer\,%
\thanks{\,Supported in part by the {\sc esprit} Long Term
Research Programme of the EU under project number 20244 ({\sc alcom-it}).  
Research carried out while this author was at the Universit\'e de
Montr\'eal.}\\ {\protect\small\sl Odense University\/}\,%
\thanks{\,Department of Mathematics and Computer Science, Odense
University, Campusvej~55, \mbox{DK--5230} \mbox{Odense~M}, Denmark.
email: {\tt u2pi$\mathchar"40$imada.ou.dk}.}  }

\date{3 December 1996}

\begin{document}

\maketitle

\begin{abstract}
We investigate the power of quantum computers when they are required to
return an answer that is guaranteed correct after a time that is
upper-bounded by a polynomial in the worst case.  In an oracle setting,
it is shown that such machines can solve problems that would take
exponential time on any classical bounded-error probabilistic computer.
\end{abstract}
\thispagestyle{empty}

\section{Introduction}
According to the modern version the Church--Turing thesis, anything that
can be computed in polynomial time on a physically realisable device can
be computed in polynomial time on a probabilistic Turing machine with
bounded error probability.  This belief has been seriously challenged by
the theory of quantum computing.  In~particular, it was shown by Peter
Shor that quantum computers can factor large numbers in polynomial
time~\cite{Shor94}, which is conjectured to be impossible for classical
devices.  However, Shor's algorithm is polynomial-time in the expected
sense: there is no upper bound on how long it will run on any given
instance if we keep being unlucky.  In~this paper, we address the
question of {\em Exact\/} Quantum Polynomial Time, which concerns the
problems that quantum computers can solve in guaranteed worst-case
polynomial time with zero error probability.  Note that this strong
requirement would make randomness useless for classical machines:
anything you can compute on a classical probabilistic computer with zero
error probability in guaranteed worst-case polynomial time can be done
in polynomial-time by a {\em deterministic\/} computer---simply run the
probabilistic algorithm with an arbitrarily fixed sequence of coin
``tosses''.

The study of Exact Quantum Polynomial Time is not new.  The very first
algorithm ever designed to demonstrate an advantage of quantum computers
over classical computers, due to Deutsch and Jozsa~\cite{DJ92}, was of
this Exact nature.  However, it solved a problem that could be solved
just as efficiently on a classical probabilistic computer, provided an
arbitrarily small (one-sided) error probability is tolerated.  Here we
demonstrate for the first time the existence of a relativized problem
that can be solved in Exact Quantum Polynomial Time, yet it would
require exponential time to obtain a correct answer with probability
significantly better than $1/2$ by any classical probabilistic (or
deterministic) algorithm.

When it comes to decision problems, the well-known classical classes
\PT{}, \ZPP{} and \BPP{}~\cite{gill} give rise to their natural quantum
counterparts \QP{}, \ZQP{} and \BQP{}, respectively.  A~decision problem
belongs to \QP{} if it can be solved by a quantum algorithm whose answer
is guaranteed correct and whose running time is guaranteed to be bounded
by some fixed polynomial.  It~is in \ZQP{} if it can be solved with zero
error probability in quantum polynomial time: the answer is still
guaranteed correct, but the running time is required to be polynomial
merely in the expected sense (for each possible input).  This
corresponds to the classical notion of Las Vegas algorithms.  Finally,
the decision problem is in \BQP{} if it returns the correct answer with
probability better than 2/3 on all inputs, after a time that is bounded
by a polynomial.  (In~this case, it makes no difference whether we
consider expected or worst-case time.)  This corresponds to the
classical notion of Monte Carlo algorithms.  Please note that the name
``\EQP{}'' has been used by different authors, sometimes to mean
\QP{}~\cite{BB94} and sometimes to mean \ZQP{}~\cite{BV}.  To~avoid
confusion, we shall refrain from using it at~all.

The following results are known in quantum complexity theory: \mbox{$\PT
\subseteq \QP$}~\cite{benioff,bennett}, \mbox{$\ZPP \subseteq \ZQP$},
\mbox{$\BPP \subseteq \BQP \subseteq \PT^{\sharpPs}$}~\cite{valiant,BV}
and \mbox{$\BQP^{\BQPs} = \BQP$}~\cite{BBBV}.  It~is believed that $\ZQP
\not\subseteq \BPP$ because Shor's quantum factorization
algorithm~\cite{Shor94} allows to recognize
\[F ~=~ \{ \langle x,y \rangle ~|~ x \mbox{~has a prime divisor
smaller than~} y \} \]
in \ZQP{}, whereas if $F \in \BPP$ then factorization can be
accomplished in polynomial \mbox{expected} time by a classical Las Vegas
algorithm.  Moreover, with appropriate oracles, it is known that
\mbox{$\QP \not\subseteq \NP$}~\cite{BB94} (and therefore \mbox{$\QP
\not\subseteq \ZPP$}~\cite{BB92}), \mbox{$\BQP \not\subseteq
\BPP$}~\cite{BV,Simon94} and \mbox{$\NP \cap \coNP \not\subseteq
\BQP$}~\cite{BBBV}.

\newpage

A major open question concerns the power of the weakest of all
polynomial-time quantum classes, \QP{}, compared to that of the
strongest of all polynomial-time classical probabilistic classes,
\BPP{}.  Could it be that $\QP \subseteq \BPP$?  Or~perhaps rather $\BPP
\subseteq \QP$?  Or~are these two classes uncomparable?  What about
relativized versions of this question?  Clearly any oracle under which
$\PT=\PSPACE$ is so that $\QP=\BPP$ as well, but what about oracles that
separate \QP{} from \BPP{}?  Even though this paper provides an oracle
under which there is a problem that can be solved in Exact Quantum
Polynomial Time but requires exponential time to be solved with
probability better than 2/3 by any classical probabilistic algorithm, we
do not solve the \QP{} versus \BPP{} question because our problem is not
a decision problem.  Unfortunately, there is no obvious way to turn our
problem into a decision problem, in the way that the Deutsch-Jozsa
algorithm (which did not concern a decision problem either) was turned
into a relativized decision problem to separate \QP{} from
\NP{}~\cite{BB94}.  Thus we leave the separation of \QP{} from \BPP{} as
an open problem.

We assume in this extended abstract that the reader is familiar with the
basic notions of quantum computing~\cite{review,CTR}.

\section{The Problem and its Quantum Solution}
We consider a computational problem inspired from Simon's
problem~\cite{Simon94}.  Let $n \geq 2$ be any given integer.  Let
$\oplus : \{0,1\}^n \times \{0,1\}^n \rightarrow \{0,1\}^n$ denote the
bitwise exclusive-or.  Define a dot product $(\cdot) : \{0,1\}^n \times
\{0,1\}^n \rightarrow \{0,1\}$ by $a \cdot b = \left( \sum_{i=1}^n a_i
b_i \right) \text{{\sf \,mod }} 2$ where $a = a_n \dots a_1$ and $b =
b_n \dots b_1$.

\begin{description}
\item[Given:] An integer $n \geq 2$ and a function $f:\{0,1\}^n
    \rightarrow \{0,1\}^{n-1}$.
\item[Promise:] There exists a nonzero element $s \in \{0,1\}^n$ such
    that for all $x,y \in \{0,1\}^n$, \mbox{$f(x)=f(y)$} if and only if
    $x=y$ or $x=y \oplus s$.
\item[Problem:] Find a nonzero element $z \in \{0,1\}^n$ such that $s
    \cdot z = 0$.
\end{description}

Simon's original problem~\cite{Simon94} is equivalent to the problem of
determining the unknown string~$s$.  Our problem is reducible to that
one, since if we know~$s$ we can easily find a nonzero element $z \in
\{0,1\}^n$ with $s \cdot z = 0$.  However, Simon's quantum algorithm
cannot be used to solve our problem because it finds $s$ in a time that
is polynomial merely in the expected sense.  There is a nice
group-theoretic interpretation for our problem, and since that
interpretation also helps simplify the notation, we shall use~it.
Hence, we reformulate the problem as follows.

Let~$\integer_2 = \{0,1\}$ denote the field of two elements.  For any
given integer $n \geq 2$, let~$G$ denote the group $\langle
\integer_2^n, \oplus \rangle$.  For any subgroup $K \subgroup G$, let
$K^\perp = \{g \in G \,|\, g \cdot k = 0 \text{ for all } k \in K\}$
denote the orthogonal subgroup.

\begin{description}
\item[Given:] An integer $n \geq 2$ and a function $f:G = \integer_2^n
    \rightarrow \{0,1\}^{n-1}$.
\item[Promise:] There exists a subgroup $H=\{0,s\}$ of order~2 such that
    $f$ is constant and distinct on each coset of~$H$.
\item[Problem:] Find a nonzero member $z$ of the orthogonal
    subgroup~$H^\perp$.
\end{description}

For each value in the image of~$f$, there are two preimages.  More
interesting and crucial for our algorithm we also have that, for each
value $y \in \{0,1\}^{n-2}$, there are exactly four values in~$G$ for
which $f$ evaluates to either $y0$ or~$y1$.  These four values form two
distinct cosets of~$H$.

Before giving the quantum algorithm for solving the problem, we state
the notation used in the following.  For any subset $A \subseteq G$, let
$\ket{A}$ denote the equally-weighted superposition
$\frac{1}{\sqrt{|A|}} \sum_{a \in A}\ket{a}$.  In particular, if $Hg$ is
a coset of~$H$, then \ket{Hg} denotes the superposition $\frac{1}{\sqrt
2}(\ket{g} + \ket{g \oplus s})$.

Let $\mathbf W_2$ denote the one-bit Walsh-Hadamard transform, $\mathbf
W_2=\frac{1}{\sqrt
2}\left(\begin{smallmatrix}1&\phantom{-}1\\1&-1\end{smallmatrix}
\right),$ and let $\mathbf W_2^n$ denote the Walsh-Hadamard transform
applied on each bit of a system of $n$~bits.  The result of applying
$\mathbf W_2^n$ to an $n$--bit register \ket{w} is the superposition
$\frac{1}{\sqrt{2^n}} \sum_{x \in \{0,1\}^n} (-1)^{w \cdot x} \ket{x}$.
Finally, let \smash{$\mathbf Z = \left(
\begin{smallmatrix}1&\phantom{-}0\\0&-1\end{smallmatrix}\right)$} 
denote the conditional sign-shift transform.  Our quantum algorithm uses
no other transforms than $\mathbf W_2^n$, $\mathbf Z$, and those needed
to evaluate the function~$f$.  The evaluation of function~$f$ is made
reversible by mapping $\ket{x}\ket{y}$ to $\ket{x}\ket{y \oplus f(x)}$.
Note that a second application of this process will reset the second
register since $\ket{x}\ket{y \oplus f(x) \oplus f(x)} =
\ket{x}\ket{y}$.  The complete quantum algorithm is as follows.

\bigskip
\noindent{\bf Orthogonal subgroup member algorithm}
\begin{enumerate}
\item Initialize the system to be in the zero-state \ket{0}\ket{0} where
   the first register is an $n$--bit register and the second is an
   $(n-1)$--bit register.  Initially, apply the transform~$\mathbf
   W_2^n$ to the first register, producing an equally weighted
   superposition of all elements in the group~$G$, $\frac{1}{\sqrt{2^n}}
   \sum_{g \in G} \ket{g}\ket{0}$.
\item Compute $f$ in quantum parallelism and store the result in the
   second register, producing $\frac{1}{\sqrt{2^n}} \sum_{g \in G}
   \ket{g}\ket{f(g)}$.
\item Measure all bits of the second register but the least significant,
   yielding
   \[ \textstyle \frac{1}{\sqrt 2} \big(\ket{Hg_1}\ket{f(g_1)} +
   \ket{Hg_2}\ket{f(g_2)}\big) \]
   for some $g_1, g_2 \in G$ with $g_1 \oplus g_2 \not\in H$.
\item Apply the conditional sign-shift transform~$\mathbf Z$ on the
   least significant bit in the second register, producing (up to an
   overall sign-shift) $\frac{1}{\sqrt 2} \big(\ket{Hg_1}\ket{f(g_1)} -
   \ket{Hg_2}\ket{f(g_2)}\big)$.
\item Compute $f$ again in
   order to reset the second register, producing $\frac{1}{\sqrt 2}
   \big(\ket{Hg_1}-\ket{Hg_2}\big) \ket{0}$.
\item Apply ~$\mathbf W_2^n$ to the first register again.
\item Measure the first register.  Let~$z^\star$ be the outcome.
\end{enumerate}

We~claim that this algorithm returns a nonzero member of the orthogonal
subgroup, that is, that $z^\star$ is a nonzero element in $H^\perp$.
Before proving this claim, we provide an example of the algorithm.

\begin{example}
Let $n=4$ and $G=\integer_2^4$.  Let the unknown string~$s$ be $0101$
and the unknown subgroup $H=\{0,s\}$.  The function~$f$ is constant on
each coset of~$H$, so we need only specify it on a transversal~$T$
for~$H$, say,
\[\begin{array}{c|*8c}\hline
T&0000&0001&0010&0011&1000&1001&1010&1011\\
f& 000& 010& 100& 110& 101& 001& 011& 111\\\hline
  \end{array}\]

After the second step of the quantum algorithm, the system is in a
superposition of all elements in the group, $\frac{1}{4}\sum_{g \in G}
\ket{g}\ket{f(g)}$.  Suppose we measure the string~$01$ in the third
step.  This projects the superposition to
\[\frac12 \bigg(\big(\ket{0001} + \ket{0100}\big)\ket{010} + 
          \big(\ket{1010} + \ket{1111}\big)\ket{011}\bigg).\] 
Applying the conditional sign-shift transform and uncomputing~$f$
produces
\[\frac12 \big(\ket{0001}+\ket{0100}-\ket{1010}-\ket{1111}\big)
   \ket{000},\]
and applying the final Walsh-Hadamard transform gives the superposition
\[\frac12 \big(\ket{0010}+\ket{1000}-\ket{0101}-\ket{1111}\big)
   \ket{000}.\]
It~can easily be verified that each of the four basis-states in this
superposition holds a nonzero member of the orthogonal
subgroup~$H^\perp$ in the first register.
\end{example}

\begin{theorem}
Let $n \geq 2$ and $G = \integer_2^n$.  Let $H \subgroup G$ be an
unknown subgroup of order~$2$.  Let~$f:G \rightarrow \{0,1\}^{n-1}$ be
any function constant and distinct on each coset of~$H$.  Then the
quantum algorithm given above finds a nonzero member of $H^\perp$ in
time polynomial in~$n$ and in the time to compute~$f$.
\end{theorem}

The theorem follows from the simple lemma below, by observing that only
nonzero members of the orthogonal subgroup can have nonzero amplitude
after completing step~6.

\begin{lemma}
Let $G=\integer_2^n$ and $H =\{0,s\} \subgroup G$.  Let $g_1, g_2 \in G$
be any two elements in~$G$ such that $g_1 \oplus g_2 \not\in H$.  Let
\[\ket{\psi} = \mathbf W_2^n 
   \bigg(\frac{1}{\sqrt 2}\big(\ket{Hg_1}-\ket{Hg_2}\big)\bigg).\]
Then, for all $x \in G$,
\begin{equation*}
\langle x | \psi \rangle = \begin{cases}
\pm \frac{1}{\sqrt{2^{n-2}}} & \text{if } x \cdot s = 0
   \text{ and } x \cdot (g_1 \oplus g_2) = 1\\
0 & \text{otherwise.}
\end{cases}
\end{equation*}
\end{lemma}

\begin{proof}
The amplitude of state \ket{x} in superposition \ket{\psi} is given by
\[\frac{1}{2\sqrt{2^n}}
  \bigg((-1)^{g_1 \cdot x} + (-1)^{(g_1 \oplus s) \cdot x} - 
  (-1)^{g_2 \cdot x} - (-1)^{(g_2 \oplus s) \cdot x}\bigg).\]
This can be factorized as
\[\frac{1}{2\sqrt{2^n}} (-1)^{g_1 \cdot x}
  \bigg(1 + (-1)^{s \cdot x}\bigg) \bigg(1 - (-1)^{(g_1 \oplus g_2) \cdot
  x}\bigg),\]
and the lemma follows.
\end{proof}

\newcommand{\prob}[1]{\mbox{\rm Prob}[#1]}
\section{Classical Lower Bound}
In this section, we prove that any classical algorithm that would try to
solve the above problem in subexponential time would have probability
exponentially close to 1/2 to give a correct answer, which is
essentially no better than guessing an answer at random.  This is
captured in the following theorem and its immediate corollary.

\begin{theorem}\label{mainthm}
Consider an integer $n \geq 2$ and pick a function $f:\{0,1\}^n
\rightarrow \{0,1\}^{n-1}$ at random according to the uniform
distribution among all functions that satisfy the promise that there
exists an $s \in \{0,1\}^n$ such that $f(x)=f(y)$ if and only if $x
\oplus y = s$ for all distinct $x$ and $y$ in $\{0,1\}^n$.  Consider an
arbitrary classical algorithm that has access to $f$ as an oracle.
Assume the algorithm makes no more than $2^{n/3}$ calls on its oracle.
Then there exists an event $\cal E$ such that (1) $\prob{{\cal E}} <
2^{-n/3}$ and, (2) If $\cal E$ does {\em not\/} occur then the
probability that the algorithm returns a nonzero $z \in \{0,1\}^n$ such
that $s \cdot z=0$ is less than $\frac{1}{2}+2^{-n/3}$.
\end{theorem}

\begin{proof}
This theorem follows directly from Lemmas~\ref{lem1} and~\ref{lem2},
which are stated and proven below.
\end{proof}

\begin{corollary}
The probability that the algorithm mentioned in Theorem~\ref{mainthm}
will return a correct answer after making no more than $2^{n/3}$ calls
on its oracle is less than $\frac{1}{2}+1/2^{(n/3)-1}$.
\end{corollary}

To establish these results, assume that the algorithm has queried its
oracle on inputs $x_1$, \mbox{$x_2$,\ldots,} $x_k$ for $x_i \in
\{0,1\}^n$, $1 \le i \le k \le 2^{n/3}$.  Without loss of generality,
assume that all the queries are distinct.  Let $y_1$, $y_2$,\ldots,
$y_k$ be the answers obtained from the oracle, i.e.~$y_i=f(x_i)$ for
each~$i$.  Define the event $\cal E$ as {\em occurring\/} if there exist
$i$ and $j$, $1 \le i < j \le k$, such that $y_i=y_j$.  Clearly, the
algorithm has discovered the secret $s$ when $\cal E$ occurs since in
that case $s=x_i \oplus x_j$.  This allows the algorithm to produce a
correct solution with certainty.  We~have to prove that $\cal E$ is very
unlikely and that, unless $\cal E$ occurs, the algorithm has so little
information that it cannot return an answer that is significantly more
probable to be correct than a random $n$--bit string.

Let $X=\{x_1,x_2,\ldots,x_k\}$ be the set of queries to the oracle and
let $Y=\{y_1,y_2,\ldots,y_k\}$ be the corresponding answers.  Let
$W=\{x_i \oplus x_j \,|\, 1 \le i < j \le k\}$ and let $m < k^2$ be the
cardinality of~$W$.  Note that $\cal E$ occurs if and only if $s \in W$
since $y_i=y_j$ if and only if $x_i \oplus x_j=s$.  If~$\cal E$ does not
occur, we say that any nonzero $n$--bit string ${\hat s} \not\in W$ is
{\em compatible\/} with the available data because it is not ruled out
as possible value for the actual unknown~$s$.  Similarly, given any
compatible $\hat s$, we say that a function ${\hat f}:\{0,1\}^n
\rightarrow \{0,1\}^{n-1}$ is {\em compatible\/} with the available data
(and with $s=\hat s$) if ${\hat f}(x_i)=y_i$ for all $i$, and if ${\hat
f}(x)={\hat f}(y)$ if and only if $x \oplus y = {\hat s}$ for all
distinct $x$ and $y$ in $\{0,1\}^n$.  The following lemma says that all
compatible values for $s$ are equally likely to be correct given the
available data, and therefore the only information available about $s$
is that it is one of the compatible values.

\begin{lemma}\label{mainlemma}
Assume $\cal E$ has not occurred.  There are exactly
$(2^n-m-1)((2^{n-1}-k)!)$ functions that are compatible with the
available data.  For each compatible string $\hat s$, exactly
$(2^{n-1}-k)!$ of those functions are also compatible with~$s=\hat s$.
\end{lemma}

\begin{proof}
Consider an arbitrary compatible~$\hat s$.  Define $X'=\{x \oplus {\hat
s} \,|\, x \in X\}$.  It~follows from the compatibility of $\hat s$ that
$X \cap X'=\emptyset$.  Let $Z=\{0,1\}^n \setminus (X \cup X')$, where
``$\setminus$'' denotes set difference.  Note that $x\in Z$ if and only
if $x \oplus \hat{s} \in Z$.  Partition $Z$ in an arbitrary way into
$Z_1 \cup Z_2$ so that $x \in Z_1$ if and only if $x \oplus \hat{s} \in
Z_2$.  The~cardinalities of $Z_1$ and $Z_2$ are $(2^n-2k)/2 =
2^{n-1}-k$.  Now let $Y' = \{0,1\}^{n-1} \setminus Y$, also a set of
cardinality $2^{n-1}-k$.  To~each bijection $h:Z_1 \rightarrow Y'$ there
corresponds a function $\hat f$ compatible with the available data and
$s=\hat s$ defined by
\[ {\hat f}(x) ~=~ \left\{
\begin{array}{ll}
y_i & \mbox{if } x=x_i \mbox{ for some } 1 \le i \le k \\
y_i & \mbox{if } x=x_i \oplus {\hat s} \mbox{ for some } 1 \le i \le k \\
h(x) & \mbox{if } x \in Z_1 \\
h(x\oplus {\hat s}) & \mbox{if } x \in Z_2.
\end{array}
\right. \] 
The conclusion follows from the facts that there are $(2^{n-1}-k)!$ such
bijections, each possible function compatible with the available data
and $s=\hat s$ is counted exactly once by this process, and there are
$2^n-m-1$ compatible choices for $\hat s$, each yielding a disjoint set
of functions compatible with the available data.
\end{proof}

\begin{lemma}\label{lem1}
Event $\cal E$ has probability of occurrence smaller than $2^{-n/3}$
provided the oracle is probed $k \le 2^{n/3}$ times.
\end{lemma}

\begin{proof}
Since all nonzero values for $s$ are equally likely {\em a~priori}, and
since event $\cal E$ occurs if and only if $s \in W$, it follows that
\[ \prob{{\cal E}} = m/(2^n-1) < k^2/2^n \le 2^{-n/3} , \]
where $m$ is the cardinality of~$W$.
\end{proof}

\begin{lemma}\label{lem2}
If event $\cal E$ does {\em not\/} occur then the probability that the
algorithm returns a nonzero $z \in \{0,1\}^n$ such that $s \cdot z=0$ is
less than $\frac{1}{2}+2^{-n/3}$, provided the oracle is probed $k \le
2^{n/3}$ times.
\end{lemma}

\begin{proof}
Assume that event $\cal E$ has not occurred after $k \le 2^{n/3}$ probes
to the oracle.  Consider an arbitrary nonzero $z \in \{0,1\}^n$.  Let
$A_z = \{ u \in \{0,1\}^n \,|\, u \cdot z = 0 \}$,
\[ B_z = \{ u \in A_z \,|\, u \neq 0^n \text{ and } u \not\in W \} \]
and let $b_z$ be the cardinality of~$B_z$.  It~is well-known that $A_z$
contains $2^{n-1}$ elements, and therefore $b_z \le 2^{n-1}-1$.  We~know
from Lemma~\ref{mainlemma} that the only knowledge about~$s$ that is
available to the algorithm is that it is nonzero and not in~$W$.
Therefore, $z$ is a correct answer if and only if $s \in B_z$, and the
optimal strategy for the algorithm is to return some $z$ that
maximizes~$b_z$.  Given that there are $2^n-1-m$ possible values
for~$s$, the probability of success (conditional to event $\cal E$
having not occurred) is
\[ \frac{b_z}{2^n-1-m} \le
   \frac{2^{n-1}-1}{2^n-1-m} <
   \frac{2^{n-1}}{2^n-k^2} \le
   \frac{2^{n-1}}{2^n-2^{2n/3}} =
   \frac{1/2}{1-2^{-n/3}} \le
   \frac{1}{2} + 2^{-n/3} 
\]
provided $n \ge 3$.  The Lemma holds also when $n=2$ since in this case
at most one question is allowed ($2^{2/3}<2$), which gives a success
probability {\em smaller\/} than 1/2 for all possible algorithms!
\end{proof}

\section{Concluding Remarks and Open Problems}
In~the quantum algorithm, we performed a partial measurement at step~3.
This step is not necessary, as we still will obtain a nonzero member of
the orthogonal subgroup if we only perform steps 1, 2 and \mbox{4--7}.
We~have, however, included it here to emphasize a group-theoretic
interpretation of the algorithm.  The algorithm (and the notion of
orthogonal subgroups) can be generalized to arbitrary finite Abelian
groups of smooth order.  The requirement of smoothness is sufficient to
be able to perform the quantum Fourier transform (step~1) and the
conditional phase-changes (step~4) exactly in polynomial time.

\begin{theorem}
Let $G$ be any Abelian group of smooth order~$m$.  Let $H \subgroup G$
be an unknown subgroup of known index~$r$, $r>1$.  Let~$f:G \rightarrow
\{0,\dots,r-1\}$ be any function constant and distinct on each coset
of~$H$.  Then there exists a quantum algorithm that finds a nonzero
member of $H^\perp$ in time polynomial in~$\log(m)$ and in the time to
compute~$f$.
\end{theorem}

An~interesting open question related to ours and Simon's algorithms is
whether~$s$ can be found in Exact Quantum Polynomial Time.  From a
complexity-theoretic point of view, an oracle separation of \QP{} and
\BPP{} is still an open question since our problem is not a decision
problem.

\end{document}